\documentclass[aps,prb,preprint,superscriptaddress,showpacs,preprintnumbers,amsmath,amssymb,10pt,twocolumn]{revtex4}
\usepackage{graphicx}
\usepackage{bm}
\usepackage[usenames, dvipsnames]{color}
\usepackage{epsfig}
\begin{document}

\title{Soft repulsive mixtures under gravity: brazil-nut effect, depletion
bubbles, boundary layering, nonequilibrium shaking}
% Force line breaks with \\

\author{Tobias Kruppa}
\author{Tim Neuhaus}
\affiliation{%
Institut f\"ur Theoretische Physik II: Weiche Materie,
Heinrich-Heine-Universit\"at
D\"usseldorf, Universit\"atsstra\ss e 1, D-40225 D\"usseldorf, Germany\\
}%
\author{Ren\'e Messina}
\affiliation{%
Institut f\"ur Theoretische Physik II: Weiche Materie,
Heinrich-Heine-Universit\"at
D\"usseldorf, Universit\"atsstra\ss e 1, D-40225 D\"usseldorf, Germany\\
}%
\affiliation{%
Laboratoire de Chimie et Physique - Approche Multi-Echelle des Milieux Complexes,
Universit\'e de Lorraine, Institut de Chimie, Physique et Mat\'eriaux (ICPM),
1 Bd. Arago, 57078 Metz - Cedex 3, France
}%
\author{Hartmut L\"owen}
\affiliation{%
Institut f\"ur Theoretische Physik II: Weiche Materie,
Heinrich-Heine-Universit\"at
D\"usseldorf, Universit\"atsstra\ss e 1, D-40225 D\"usseldorf, Germany\\
}%

\date{\today}% It is always \today, today,
             %  but any date may be explicitly specified

\begin{abstract}

A binary mixture of particles interacting via long-ranged repulsive forces
is studied in gravity by computer simulation and theory. The
more repulsive $A$-particles create a depletion zone of less repulsive
$B$-particles around them
reminiscent to a bubble. Applying Archimedes' principle effectively to this
bubble, an $A$-particle can be lifted in a fluid background of $B$-particles.
This "depletion bubble" mechanism explains and predicts a brazil
nut effect where the heavier $A$-particles float on top of the lighter
$B$-particles.
It also implies an effective attraction of an $A$-particle towards a hard
container bottom wall
which leads to boundary layering of $A$-particles. Additionally, we have
studied
a periodic inversion of gravity causing
perpetuous mutual penetration of the mixture in a slit geometry.
In this nonequilibrium case of time-dependent gravity, the boundary
layering persists.
Our results are based on computer simulations and
density functional theory
of a two-dimensional binary mixture of colloidal repulsive dipoles.
The predicted effects also  occur for other
long-ranged repulsive interactions and in three spatial dimensions. They
are therefore verifiable in settling  experiments on dipolar or charged
colloidal mixtures  as well
as in charged granulates and dusty plasmas.
\end{abstract}

\maketitle

\section{Introduction}

Gravity or centrifugation is commonly used to sort and separate different
particles
out of a mixture \cite{Philipse,Pine} but the underlying microscopic
(i.e.\ particle resolved)
processes of mixing and demixing under settling are still debated
\cite{Rasa,Louis_Padding}.
It is known for long time that shaken or vibrated granular mixtures can
exhibit the "brazil-nut effect",
namely that the heavier particles are on top of the lighter ones
\cite{williams1976,rosato1987}. The details and parameter combinations for
the brazil-nut effect
to occur are still discussed
\cite{shinbrot1998,hong2001,both2002,breu2003,Godoy,Garzo}.
The brazil-nut effect may even contradict
  Archimedes' law, which governs the equilibrium density profiles of
molecular mixtures and colloidal solutions by the buoyancy principle.

Colloidal mixtures are valuable model systems to explore gravity effects
on the particle scale
\cite{Wysocki,HL_SM,Brambilla,Serrano,biben1994,milinkovic2011} both in
equilibrium and nonequilibrium.
Sediments of binary charged mixtures are commonly used to determine the
phase behavior \cite{Lorenz,Okubo,Vermolen,Weeks}.
In highly deionized charged colloidal mixtures, the sediment
\cite{Esztermann,Zwanikken} can split into separated layers
due to counterion lifting, a phenomenon referred to as "colloidal brazil
nut effect". The separation of
binary hard-sphere mixtures was explored using the equation of state and
separation of the two species was predicted in line with Archimedes'
principle \cite{Biben_binary,Wasan1,Wasan2}.
Recent experimental real-space studies on soft repulsive colloidal
mixtures in three dimensions
\cite{Serrano} clearly showed that the buoyancy principle is not violated.
Colloid-polymer mixtures are known to phase-separate under gravity in
equilibrium \cite{floating1,floating2}.

Nonequilibrium studies, on the other hand, include the dynamics of the
settling process
on the particle scale \cite{Schmidt_Dzubiella,Wysocki} after quickly
turning the sample upside down, the enforcement of crystal growth
on a patterned template under gravity
\cite{Hoogenboom2,Ramsteiner}, spatially varying temperature fields
\cite{Puri} and novel zone formation in sedimenting colloidal mixtures
\cite{Leocmach}.
Interestingly, for colloidal mixtures, there are only few studies where
gravity is changed periodically in time
\cite{Smorodin} which may be considered to be  the colloidal analogue of
granulate shaking.

In this paper, we consider a binary mixture of particles interacting via
long-ranged repulsive forces
in gravity. We use Monte Carlo and Brownian dynamics computer simulation
and mean-field
density functional theory to predict the equilibrium density profiles and
the nonequilibrium response
of the system to oscillatory gravity. The more repulsive particles are
referred to as $A$-particles while the less repulsive particles are the
$B$-particles.
$A$-particles create
a depletion zone of small particles around them
reminiscent to a bubble. Applying Archimedes' principle effectively to this
bubble, an $A$-particle can be lifted in a fluid background of $B$-particles.
This "depletion bubble" mechanism results in  a brazil
nut effect where the heavier $A$-particles float on top of the lighter
$B$-particles.
It also implies an effective attraction of an $A$-particle towards a hard
container bottom wall
which produces a boundary  layering of the $A$-particles.
If the direction of gravity is periodically inverted causing
perpetuous mutual penetration of the mixture in a slit geometry, a similar
stable layering
emerges as a nonequilibrium phenomenon.
We emphasize that our effects do not occur for short-ranged interactions
like for hard sphere mixtures \cite{Biben_binary}. This is only the case
when the (effective)
hard sphere interaction
diameter differs largely from that of the actual particle size. Therefore, it
 is the softness  of the repulsion which is relevant
here.

Our results are obtained for
a two-dimensional binary mixture of colloidal repulsive dipoles.
Therefore, our simulation results can be verified in real-space
microscopy experiments of two-dimensional superparamagnetic particles
\cite{Naegele,Konig,Ebert_EPJE_2008,Mazoyer,Keim2,Keim3}
see also \cite{Ling,Wang} for alternative set-ups. An external magnetic
field induces repulsive dipole forces.
The gravity can either be realized by tilting the droplets or by applying a
laser light pressure on the sample \cite{Jenkins}.  The predicted effects
also  occur for other
long-ranged repulsive interactions and in three spatial dimensions. They
are therefore verifiable in settling  experiments on dipolar or charged
colloidal mixtures \cite{Serrano} as well
as in charged granulates \cite{gran} and dusty plasmas
\cite{dust,morfill2009}.

The paper is organized as follows: in section II, we describe the model
and the simulation technique applied.
In section III, we discuss the density functional theory. Results are
presented in sections IV and V and we conclude  in section VI.

\section{Model and simulation technique}

The system consists of a suspension of two species of point-like
super-paramagnetic colloidal particles denoted as $A$ and $B$, which are
confined to a two-dimensional planar interface.
These particles are characterized  by different
magnetic dipole moments $M_A$ and $M_B$, where
%
%%%%%%%%%%%%%%%%
\begin{equation}
\label{eq:M}
M=M_B/M_A
\end{equation}
%%%%%%%%%%%%%%%%
is the dipole-strength ratio.
The dipoles are induced by an external magnetic field $H$ according to
$M_i = \chi_i H$
($i=A,B$), where $\chi_i$ denotes the magnetic susceptibility. The
magnetic field is applied perpendicular
to the two-dimensional interface containing the particles.
In the following, the  dipole-strength ratio $M$ is fixed to 0.1,
corresponding to recent experimental
samples \cite{Ebert_EPJE_2008,Mazoyer,assoud2009}.
The relative composition $X=N_B/(N_A+N_B)$ of $B$ particles is fixed at
$50\%$; hence we are considering an equimolar mixture.
The particles are exposed to an external potential $V_{ext,i}({\bm{r}})$
which is a combination of gravity
and the hard bottom wall and is given by
%
%%%%%%%%%%%%%%%%
\begin{equation}
\label{eq:V_ext}
V_{ext,i}(\bm{r}) =
\begin{cases}
  m_i g y  & \text{for} \quad y \geq 0\\
  \infty & \text{otherwise}
\end{cases}
\end{equation}
%%%%%%%%%%%%%%%%
%
Here, $m_{i}$ is the buoyant mass of particle species $i$ $(i=A,B)$.
Gravity acts along the $-y$
direction. We characterize the mass ratio by the dimensionsless parameter
%%%%%%%%%%%%%%%%
\begin{equation}
\label{eq:m}
m=m_B/m_A \quad .
\end{equation}
%%%%%%%%%%%%%%%%
The particles interact via  a repulsive  pair potential of two parallel
dipoles of the form
%
%%%%%%%%%%%%%%%%
\begin{equation}
\label{eq:u_r}
u_{ij}(r)=\frac{\mu_0}{4\pi}M_i M_j /r^3 = \frac{\mu_0}{4\pi}\chi_i \chi_j
H^2/r^3
\quad  (i,j=A,B),
\end{equation}
%%%%%%%%%%%%%%%%
where $r$ denotes the distance between two particles in the plane.
For this inverse power potential, at fixed composition $X$ and
susceptibility ratio
$\chi_B/\chi_A$, all static quantities
depend solely \cite{Hansen_MacDonald_book} on a dimensionless interaction
strength
(or coupling constant)
%
%%%%%%%%%%%%%%%%
\begin{equation}
\Gamma=\frac{\mu_0}{4\pi}\frac{\chi_A^2H^2}{k_BTl_A^3} \quad ,
\end{equation}
%%%%%%%%%%%%%%%%
%
where $k_BT$ is the thermal energy
and  $l_A=k_BT/(m_Ag)$ the gravitational length of $A$ particles, which we
employ as a unit of length.

For time-dependent gravity ("shaking") we consider the external potential
%%%%%%%%%%%%%%%%
\begin{equation}
\label{eq:V_ext_{t}}
V_{ext,i}(\bm{r},t) =
\begin{cases}
  m_i g(t) y  & \text{for} \quad 0 \leq y \leq L_y\\
  \infty & \text{otherwise}
\end{cases}
\end{equation}
%%%%%%%%%%%%%%%%
which embodies a time-dependent gravity strength $g(t)$ in a finite slit
of width $L_{y}$.
This is conveniently
modelled to be a stepwise constant function of time:
%%%%%%%%%%%%%%%%
\begin{align}
\label{eq:g_t}
&g(t) =
\begin{cases}
 g & \text{for} \quad n -1 < t / T_0 \leq n - \vartheta\\
 - g & \text{for} \quad n - \vartheta < t / T_0 \leq n 
\end{cases}\\[0.1cm]
&\qquad \qquad n = 1,2,.. \nonumber
\end{align}
%%%%%%%%%%%%%%%%
introducing a time period $T_{0}$ and a ''swap fraction'' $0 \leq \vartheta \leq 1$.
Note that for the time-dependent gravity we confine the system to a slit of
vertical width $L_{y}$ in order to keep the external potential bounded
from below for all times.

The particle dynamics is assumed to be Brownian. Hydrodynamic interactions
are neglected.
The Brownian time scale is set by the short-time diffusion constant $D_A$
 of the $A$-particles. Knowing that this diffusion constant scales with
the inverse of the radius of 
a particle, $D_B$ was chosen such that
$D_B/D_A=1.61$ corresponding to the physical diameter ratio
 of the experimental samples \cite{Ebert_EPJE_2008,Mazoyer,assoud2009}.

We  perform  standard nonequilibrium Brownian dynamics (BD) and Monte
Carlo (MC) computer
simulations \cite{allen1989,stirner2005,Frenkel2001} in the canonical
ensemble
where the particle numbers $N_{i}$ $(i=A,B)$, the temperature $T$ and the
area $\mathbf{A}$ is fixed.
In the static (equilibrium) case, where the external potential is not
time-dependent, the one-particle density field has been calculated via MC
simulations of $N_A=600$  $A$-particles and $N_B=600$
$B$-particles, which were placed in a finite rectangular box $L_x \times
L_y$ of area $\mathbf{A} = (30 \times 120) (l_A)^2$. In the nonequilibrium case,
where gravity is time-dependent, we obtained the one-particle density
field by performing BD simulations of $N_A=N_B=300$ particles in a
rectangular box of size $\mathbf{A} = (15 \times 60) (l_A)^2$. In both cases, the
simulation box features periodic boundary conditions in $x$-direction, an
aspect ratio $L_{x}/L_{y} = 1/4$ and a gravitational load $N_A / L_x = 20
/l_A$. The coupling constant $\Gamma$ is fixed to 10. A finite time step  $\delta t=10^{-4}\tau$ was used in the BD
simulations, where $\tau=l_A^2/D_A$. We denote the density profiles as
$\rho_i^{\text{eq}}(\bm{r})$ $(i=A,B)$ in the static case and $\rho_i(\bm{r},t)$
$(i=A,B)$ in the dynamical case.

\section{Density functional theory}

Within density functional theory (DFT)
 the grand canonical free energy
$\Omega(T,\mu_A,\mu_B,[\rho_A({\bm{r}}),\rho_B({\bm{r}})])$ depending on
the temperature $T$ and  the chemical potentials $\mu_A, \mu_B$ is
minimized with respect to the local partial one-particle
densities $\rho_A({\bm{r}})$ and $\rho_B({\bm{r}})$.
This functional can be split according
to \cite{mermin1965,evans2002,lowen1994}, so that in two spatial dimensions
we obtain
\begin{align}
 &\Omega\left(T,\mu_A,\mu_B,[\rho_A({\bm{r}}),\rho_B({\bm{r}})]\right)
=\mathcal{F}_{\rm id}\left([\rho_A({\bm{r}}),\rho_B({\bm{r}})]\right)+ \nonumber \\
&\mathcal{F}_{\rm exc}\left([\rho_A({\bm{r}}),\rho_B({\bm{r}})]\right)
		+\sum \limits_{i=\rm A,B} \int \text{d}^2r\rho_i({\bm{r}})\left[V_{{\rm
ext},i}({\bm{r}})-\mu_i\right],
 \label{eq:omega_dft}
\end{align}
where the first term is the free energy of an ideal gas
\begin{equation}
 \mathcal{F}_{\rm id}=k_BT \sum \limits_{i=\rm A,B} \int
\text{d}^2r\rho_i({\bm{r}})\left[\ln(\Lambda_i^2\rho_i({\bm r}))-1\right]
\end{equation}
including the (irrelevant) thermal wavelength $\Lambda_i$ of particles of
species $i$ $(i=A,B)$.
As already introduced above, $V_{{\rm ext},i}({\bm r})$ is the static
external potential
acting on particle species $i$. The only unknown part is the excess free
energy functional $\mathcal{F}_{\rm exc}$, resulting from the
inter-particle interactions. In order to approximate $\mathcal{F}_{\rm
exc}$,
we use a simple Onsager functional \cite{Onsager1949,Wensink2008}
\begin{equation}
 \mathcal{F}_{\rm exc}=\frac{k_BT}{2}\int \text{d}^2r\int \text{d}^2r'
f_{ij}(|
{\bm r}-{\bm r}'|)\rho_i({\bm r})\rho_j({\bm r'}),
\end{equation}
consisting of the Mayer $f$-function
\begin{equation}
 f_{ij}({\rm r})=1-e^{-\beta u_{ij}(r)},
\end{equation}
with the interaction potential $u_{ij}(r)$ from equation ~\eqref{eq:u_r} and the
inverse temperature $\beta=(k_BT)^{-1}$.
The Onsager approximation is valid at low densities reproducing the second
virial
coefficient of the bulk fluid equation of state correctly but is expected
to break down at higher densities.
Unfortunately, unlike for hard-core interactions
\cite{rosenfeld1997,oettel2010},  no alternative
approximation working at higher densities is known for soft
inverse-power-law potentials \cite{footnote1}.
However, we expect the general trends to be captured by the theory but not
the details of molecular layering.

The equilibrium density profiles $\rho_i^{\text{eq}}({\bm r})$ are obtained from
the minimization condition
\begin{equation}
 \left.\frac{\delta \Omega[\rho_A({\bm r}),\rho_B({\bm r})]}{\delta
\rho_i({\bm r})}\right|_{\rho_i({\bm r})=\rho_i^{\text{eq}}({\bm r})}=0.
\end{equation}
It is important to note that DFT is typically formulated in the
grand-canonical ensemble
where the chemical potentials $\mu_A, \mu_B$ are fixed instead of the
particle numbers, while the simulations
are performed in the canonical ensemble. We have therefore considered the
chemical potentials $\mu_A, \mu_B$
as Lagrange multipliers which fix the total line density perpendicular to
gravity and matched them such
that this line density coincides with that prescribed in the simulations.

If gravity gets time-dependent, there is a dynamical generalization of DFT
appropriate for Brownian systems
which can be derived in various ways
\cite{Marconi1999,Archer2004,espanol2009} from the exact
Smoluchowski equation via an adiabatic approximation. Within this
dynamical density
functional theory (DDFT), the time-dependent density fields obey the
generalized diffusion equation
\begin{equation}
 \frac{\partial \rho_i({\bm r},t)}{\partial t}=\beta D_{i}
\nabla\cdot\left(\rho_i({\bm r},t)\nabla \frac{\delta \Omega[\rho_A({\bm
r},t),\rho_B({\bm r},t)]}{\delta \rho_i({\bm r},t)}\right),
\end{equation}
with a diffusion constant $D_i$ corresponding to particle species $i = A,B$.
It is important to note that this equation conserves the total density,
i.e.\ provided the
starting density profiles are matched to that in the canonical ensemble,
the time evolution
given by the DDFT equation is canonical. Therefore,
the results obtained from DDFT can directly be compared to our BD
simulations. In our case of an external potential which depends only on
the $y$-coordinate, we consider only density profiles which are
independent of $x$. This is justified far away from surface freezing
\cite{heni2000}.

\section{Results in Equilibrium}

\subsection{Colloidal brazil-nut effect}

\begin{figure}[ht]
\centering
\includegraphics[width = 1.0\linewidth]{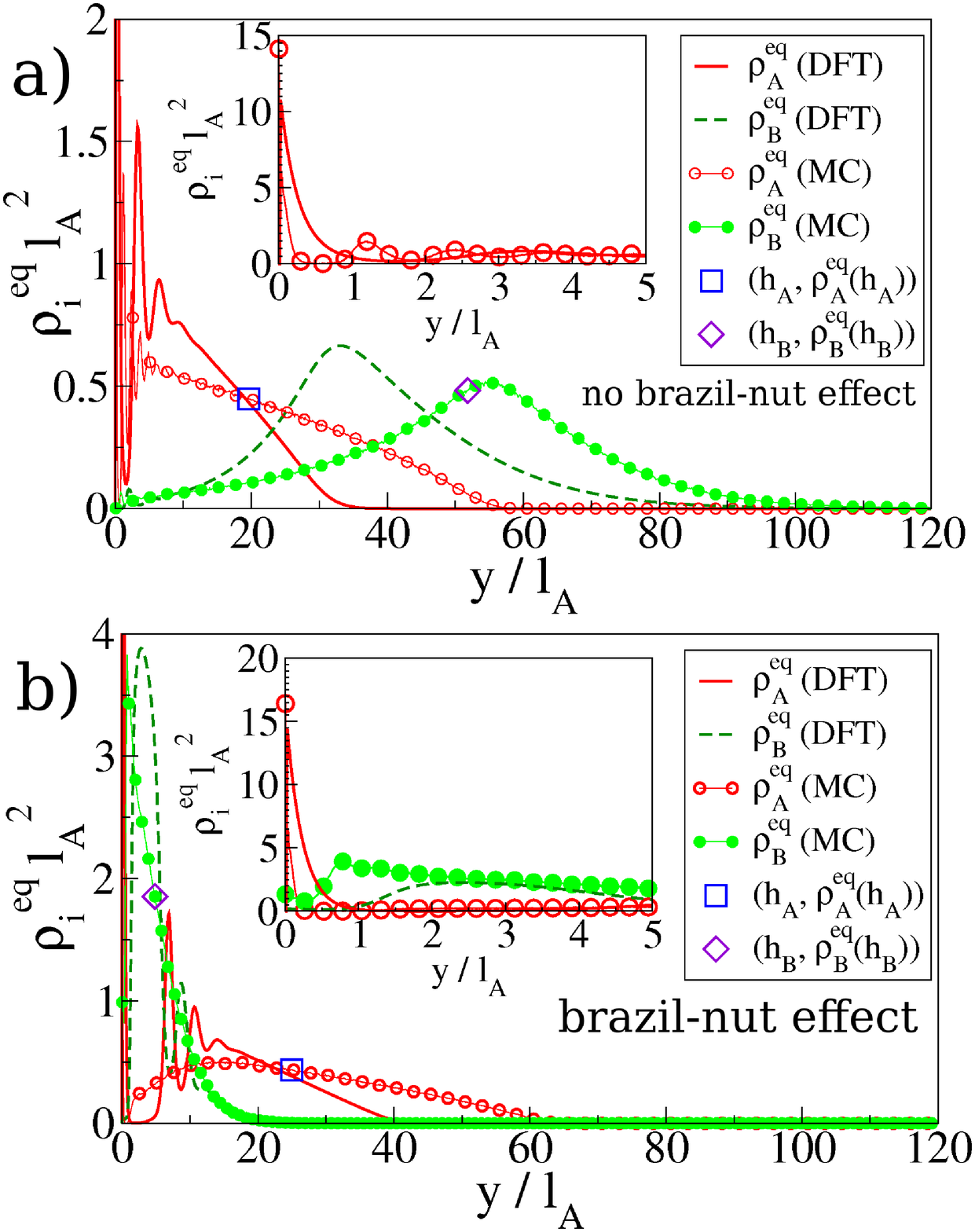}
\caption{(Color online) One-particle density profiles $\rho_i^{\text{eq}}({\bm
r})$, $i = A, B$ obtained from DFT and MC simulation. Mean sedimentation
heights $h_i$, $i = A,B$ from Eqn. (\ref{eq:h}) are indicated by
highlighted points for MC simulation data. a) $M = 0.1, m = 0.1$ (no brazil-nut effect).
b) $M = 0.1, m = 0.5$ (brazil-nut effect). The insets enlarge the
behavior at small wall distances where a strong peak of the $A$-particles
occurs.}
\label{fig:rho_010}
\end{figure}

We performed MC simulations for various mass ratios $0 \leq m \leq 1 $
and dipolar ratios $0 \leq M \leq 1$. Thereby, we choose $A$ as the
heavier and stronger coupled species
\cite{footnote2}.
An example for the partial density profiles $\rho_i^{\text{eq}}({\bm r})$ $(i=A,B)$
is given in Fig. \ref{fig:rho_010} where at
fixed $M=0.1$ two different mass ratios $m=0.1, 0.5$ are considered.
Interestingly, the MC simulation data show quite distinct qualitative
behavior for these two cases. In Fig. \ref{fig:rho_010}a ($m=0.1$)
the lighter $B$-particles are on top of the heavier $A$-particles as
expected, while in Fig. \ref{fig:rho_010}b ($m=0.5$)
the  behavior is reversed: here, the heavier $A$-particles are on top of
the lighter $B$-particles.
At first glance, this opposite trend is counterintuitive. We call it - in
some analogy to granulate matter -
 (colloidal) {\it brazil-nut effect}.\\
DFT data are also included in Fig. \ref{fig:rho_010}. In fact,
static density functional theory can basically reproduce
 the  partial density profiles $\rho_i^{\text{eq}}({\bm r})$ $(i=A,B)$ though
the comparison is not quantitative
since the functional  is approximated by a  low-density expression.
In fact, as compared to the simulation data, the DFT results for the layering spacings  are too large
but the contact density of $A$-particles at the bottom wall are well reproduced in DFT
(see the insets shown in Fig. \ref{fig:rho_010}).
Of course, one should bear in mind
that there is no fit parameter involved in the comparison.\\
\\
In order to quantify the colloidal brazil-nut effect, we follow the
criterion proposed
in Ref.\  \cite{Esztermann}. We define
 averaged heights $h_i$, $i = A,B$ by taking the first moment
of the partial density fields as
\begin{equation}
\label{eq:h}
h_i = \frac{\int_0^\infty y \rho_i^{\text{eq}}(y) \mathrm{d}y}{\int_0^\infty \rho_i^{\text{eq}}(y)
\mathrm{d}y}, \qquad i = A,B
\end{equation}
In Fig. \ref{fig:rho_010}, the location of the corresponding heights are
indicated by a large symbol.
The brazil-nut effect is then defined by the condition
\begin{equation}
\label{eq:h_bnut}
h_A < h_B \ ,
\end{equation}
which means that on average, the heavier $A$-particles are on top of the
lighter $B$-particles.

\begin{figure}
 \centering
 \includegraphics[width = 1.0\linewidth]{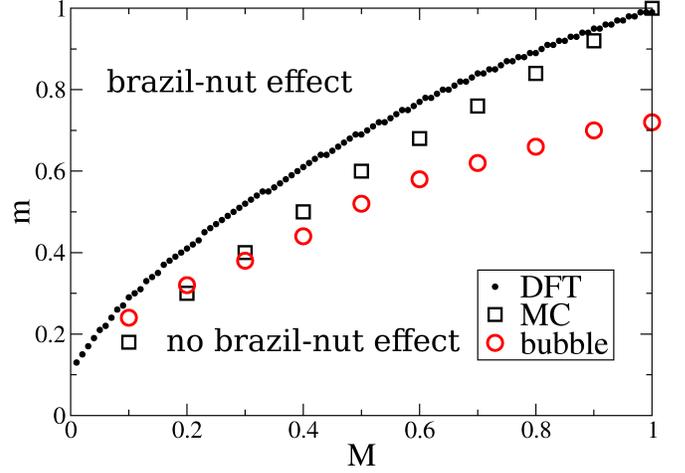}
 \caption{(Color online) Separation line between the occurrence of the
colloidal brazil-nut effect and the absence of this effect in the
parameter space of dipolar asymmetry $M$ and mass asymmetry $m$. Monte
Carlo simulation data (contoured white squares), density functional data
(full black circles) and the transition line implied by the bubble
condition (\ref{eq:buoyancy_redux}) (contoured circles) are shown.}
 \label{fig:bnut}
\end{figure}

Within the full parameter range $0 \leq m \leq 1$,  $0 \leq M \leq 1$, the
region separating
the brazil-nut effect from the ordinary behavior (no brazil-nut effect)
is shown in Fig. \ref{fig:bnut}.
MC computer simulation data for the separation line are given by square
symbols. These results
were obtained by systematically scanning the parameter space. The brazil-nut
effect occurs preferentially for strong dipolar asymmetry and is favoured
if the two masses do not differ much.
DFT results for the phase boundary are also included in Fig. \ref{fig:bnut}
and are in good agreement with the simulation data predicting the same
trends and the same
slope of the separation line in the $M$-$m$ parameter space.

\subsection{Depletion bubble picture}
\begin{figure}
 \centering
 \includegraphics[width = 0.8\linewidth]{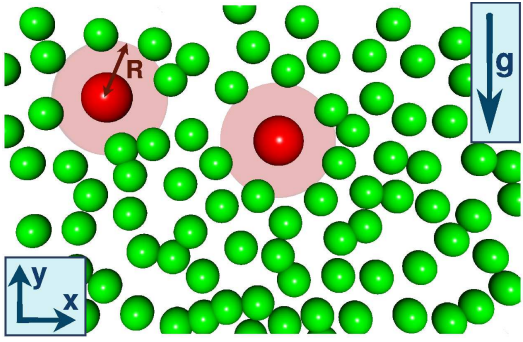}
 \caption{(Color online) Simulation snapshot depicting $A$-particles
(large red spheres) embedded in the fluid of $B$-particles (smaller green
spheres) in the bottom region of the sample. Notably, $A$-particles are
surrounded by a circular depletion zone reminiscent of a bubble (area
indicated in faint red). Particles are displayed as spheres of finite
radii for clarity only.}
 \label{fig:bubble}
\end{figure}

We now put forward an intuitive picture for the mechanism behind the
colloidal brazil-nut effect
which also provides a very simple theory for the separation line. This
picture is based on the observation
that when surrounded by a fluid of $B$-particles, a single $A$-particle 
creates a circular void-like
space around it which is free of $B$-particles due to the strong repulsion
between $A$- and $B$-particles,
see the highlighted area in Fig. \ref{fig:bubble}. This {\it depletion
bubble} is firmly attached
to the $A$-particle. Applying the buoyancy criterion to the whole bubble,
the effective weight per area of the $A$-particles is strongly reduced.
Assuming a homogeneous fluid density $\bar{\rho}_B$
surrounding a bubble of radius $R$ around an $A$-particle, the brazil-nut
condition
for lifting the $A$-particle can be stated as a buoyancy criterion
 \begin{equation}
\label{eq:buoyancy}
\frac{m_A}{\pi R^2} < m_B\bar{\rho}_B
\end{equation}
We estimate the radius $R$ by a low-density argument, where the density
profile of $B$-particles around
an $A$-particle fixed at the origin is $\bar{\rho}_B \exp\left(-
u_{AB}(r)/k_{B}T \right)=\bar{\rho}_B \exp\left(- M \Gamma l_{A}^{3}/
r^3\right)$ such that a
reasonable assumption for the radius $R$ of the depletion zone is given by
the separation where the $AB$-interaction energy equals $k_{B}T$, i.e.\
\begin{equation}
\label{eq:Rcav}
R= l_{A} (M\Gamma)^{1/3} \ .
\end{equation}
Upon insertion in Eqn. (\ref{eq:buoyancy}), this yields the brazil-nut
condition
\begin{equation}
\label{eq:buoyancy_redux}
m > \left(\pi \left(M \Gamma \right)^{\frac{2}{3}} l_A^2 \bar{\rho}_B
\right)^{-1}
\end{equation}
Thereby, a simple estimate for the separation line is provided. The only
unknown parameter entering
in Eqn. (\ref{eq:buoyancy_redux}) is the averaged density  $\bar{\rho}_B$.
We have used simulation data to determine $\bar{\rho}_B$ as the effective
density at a distance $2R$:
\begin{equation}
\label{eq:rho_B}
 \bar{\rho}_B = \rho_B^{\text{eq}}(2R)
\end{equation}
The resulting separation line is included in  Fig. \ref{fig:bnut}. Despite
its simplicity, the depletion bubble picture describes
 the simulation data pretty well.
% weitere Diskussion?
Clearly, the depletion bubble is induced by the soft long-ranged repulsion
and is therefore missing for
pure hard sphere mixtures where neighbouring particles are at contact.
However, when the interaction is mapped onto a substitute
interaction core with an effective diameter \cite{Roux_JPCM}, all the
traditional sedimentation  is qualitatively contained in this effective
hard sphere mixture.

\subsection{Boundary layering and effective interaction between an
$A$-particle and the bottom wall}
\begin{figure}
 \centering
 \includegraphics[width = 0.8\linewidth]{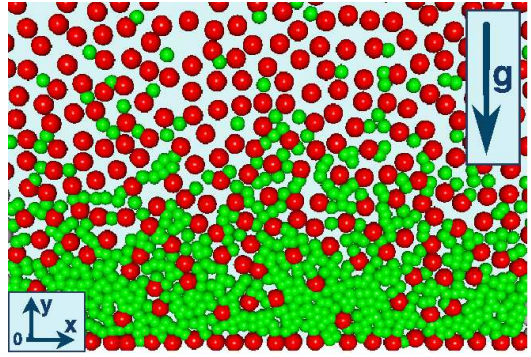}
 \caption{(Color online) Simulation snapshot for $M =0.1$, $m=0.5$ showing
the marked-off bottom layer of heavy $A$-particles (large red spheres) at
$y = 0$ beneath the fluid of light $B$-particles (small green spheres).
The arrow indicates direction of gravity, $-y$.}
 \label{fig:bottom}
\end{figure}

\begin{figure}
 \centering
 \includegraphics[width = 1.0\linewidth]{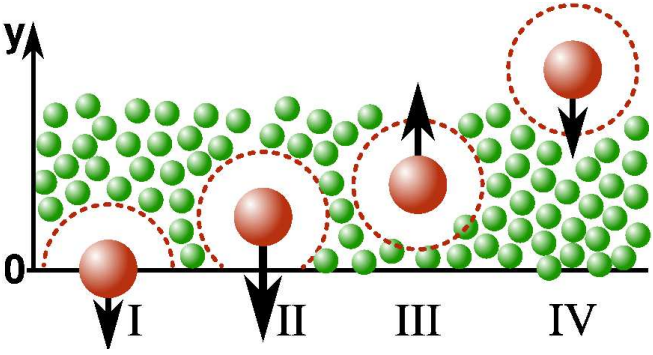}
 \caption{(Color online) Schematic illustration of the metastable trapping
of $A$-particles at $y = 0$, leading to the formation of a boundary
layer. $A$-particles are represented by large red spheres, while
$B$-particles are depicted as smaller green spheres. The solid line
indicates $y = 0$, while the orientation is analogous to Fig.
\ref{fig:bottom}. The depletion zone surrounding $A$-particles is
indicated by a dashed outline. }
 \label{fig:bottom_jump}
\end{figure}

We finally discuss the implication of the depletion bubble on the layering
of $A$-particles
close to the hard bottom wall of the confining container (at $y=0$), see
again the insets of
Fig. \ref{fig:rho_010}. The strong layering
 is clearly demonstrated by an actual simulation snapshot shown in
Fig. \ref{fig:bottom}.
If a single $A$-particle is fixed at a given distance from the bottom wall,
its depletion bubble is reduced since the void space is cut by the hard
wall, see the sketch
in Fig. \ref{fig:bottom_jump}. Note that the $A$-particle is point-like
so that in principle, it can approach the wall very closely. If the
$A$-particle is close to the wall,
the void space is half of  the full circle in the bulk (situations I and
III in Fig. \ref{fig:bottom_jump}). If the height $y$ of the $A$-particles
increases, the depletion bubble area $A(y)$ grows. Assuming a constant
depletion bubble radius $R$, $A(y)$ is given analytically as
\begin{equation}
\label{eq:A_y}
 A(y) = R^2
\begin{cases}
 \pi - \arccos(\frac{y}{R}) + \frac{y}{R}\sqrt{1 -
\left(\frac{y}{R}\right)^2} &\quad \text{for } y \leq R\\
 \pi &\quad \text{otherwise.}
\end{cases}
\end{equation}
The growing bubble size  causes two opposing effects: first, in order to
increase
the depletion bubble
area, work against the osmotic pressure $\bar{p}_B$ of the fluid
$B$-particles is necessary. Assuming that $\bar{p}_B $ is constant
in the small height regime, this work equals $\bar{p}_B (A(y)-A(0))$
and gives rise to an effective {\it attraction\/} of an $A$-particle close
to the wall (situation II
in Fig. \ref{fig:bottom_jump}). In fact, this attraction is similar to the
depletion attraction in the ordinary Asakura-Oosawa-Vrij model
of  colloid-polymer mixtures near a hard wall \cite{Dijkstra,Brader}
although there is a finite (physical) colloidal diameter in this model.

The second effect resulting from the increasing bubble size $A(y)$
is a change in the effective buoyancy. The effective buoyant force is
given by $-m_A g + \bar{\rho}_Bm_BgA(y)$. If the bubble containing the
$A$-particle is lighter than the surrounding $B$-fluid, this term is
repulsive with respect to the wall and therefore opposed to the depletion
force.

Combining these two effects, we gain the following analytical expression
for the depletion potential $V(y)$ between a single $A$-particle and the
wall:
\begin{equation}
\label{eq:V_theo}
V(y) =
\begin{cases}
 \bar{p}_B (A(y) -\pi R^2 / 2 ) + m_A gy \\
	- \bar{\rho}_Bm_Bg\int_0^y\mathrm{d}y^{\prime}A(y^{\prime}) &\text{for } y \geq 0\\[0.1cm]
 \infty &\text{for } y < 0
\end{cases}
\end{equation}

where the integral can be calculated as
\begin{align}
\label{eq:anal}
\int_0^y\mathrm{d}y^{\prime}A(y^{\prime}) = & \pi R^2 y - R^2 y\arccos\left(\frac{y}{R}\right) \nonumber \\& + R^3\sqrt{1 -
\frac{y^2}{R^2}} - \frac{R^3}{3}\left(1 -\frac{y^2}{R^2}\right)^{\frac{3}{2}}
\end{align}

This expression requires $\bar{p}_B$  as an input parameter. In order to
evaluate $V(y)$, we have determined $\bar{p}_B$  via a bulk reference
simulation of a pure $B$-system (in the absence of gravity) at a
prescribed number density $\bar{\rho}_B$ by using the virial expression
\cite{Loewen1992}.

We further checked a posteriori whether the radius $R$ of the depletion
bubble is consistent with that obtained from a radially averaged density profile of a bulk
$B$-fluid around a single $A$-particle fixed at the origin. This ''renormalized'' radius
$R^{\prime}$ can be estimated to be the position of the first inflection point 
in this density profile of $B$-particles (not shown here). Actually, we
find values for $R^{\prime}$ which are a bit smaller than those given by the 
estimate (\ref{eq:Rcav}) which only works at low $\bar{\rho}_B$.

More rigorously, we can define the effective interaction between the wall
and an $A$-particle under the presence of the inhomogeneous distribution
of $B$-particles by a potential of mean-force
\cite{mcmillan1945,Hansen_Loewen}. For a given altitude $y$, the effective
interaction potential $V_{\text{eff}}(y)$ is given by
\begin{equation}
\label{eq:v_eff}
V_{\text{eff}}(y) = -\int_0^y \ \langle F_A (y) \rangle \, \mathrm{d}y \ ,
\end{equation}
where $\langle F_A(y) \rangle$ is the canonically averaged total force on
the $A$-particle in the presence of the $B$-particles (which by symmetry
points in the $y$-direction). $F_A(y)$ also contains the trivial direct
part $-m_A g$ from gravity.
\\\\

\begin{figure}
 \centering
 \includegraphics[width = 1.0\linewidth]{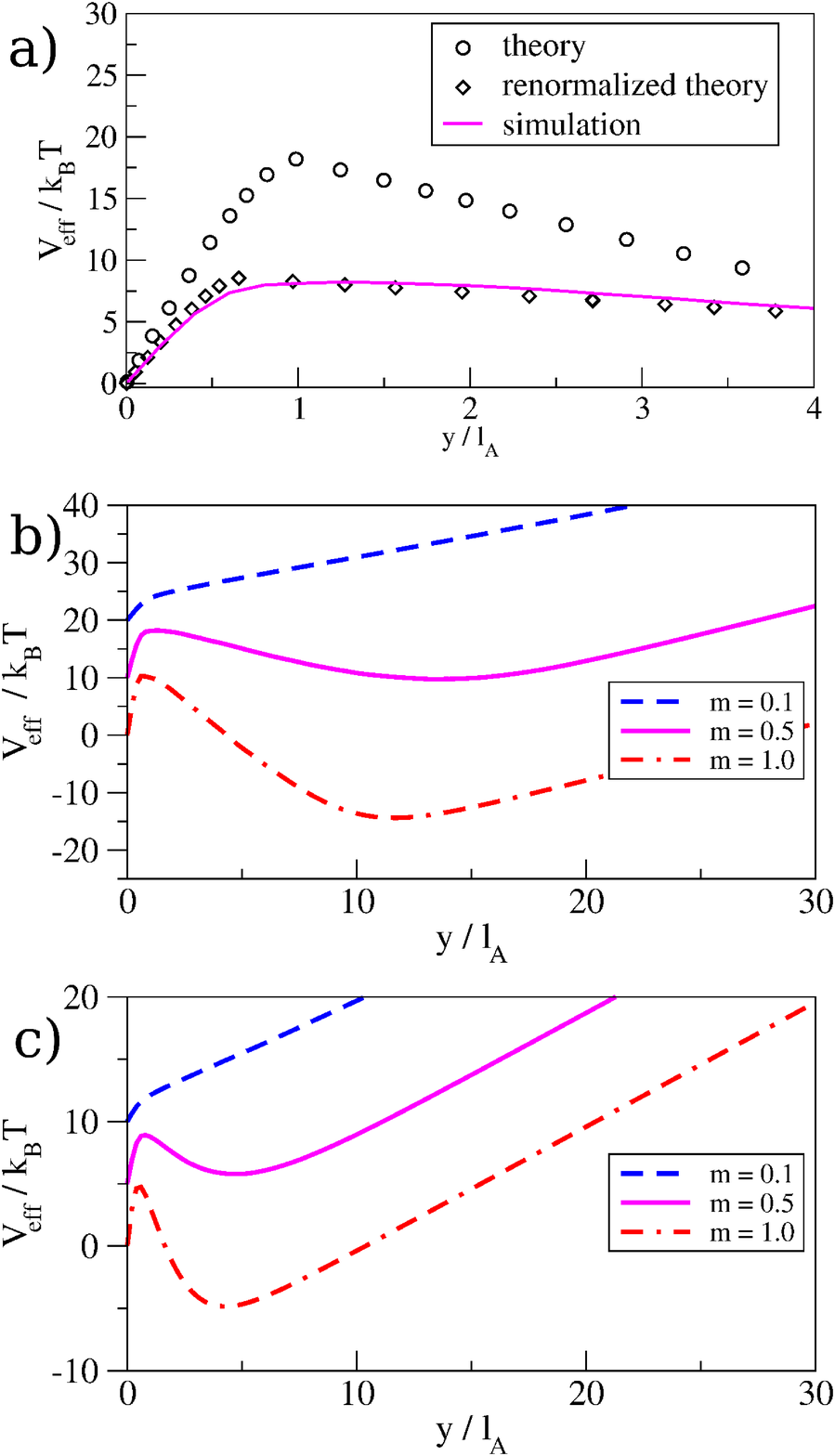}
 \caption{(Color online) Particle-wall depletion potential $V_{\text{eff}}$ 
from Eqn. (\ref{eq:v_eff}) explored by a
single $A$-particle within a fluid of $B$-particles near the system
boundary at $y = 0$ versus reduced height $y/l_A$. Panel a) shows the
potential barrier peak at close distances and includes the theoretically
predicted depletion potential $V(y)$ from Eqn. (\ref{eq:V_theo}) for
parameters $M=0.1, m=0.5$ and two different depletion bubble radii: $R$
(black circles), $R^{\prime}$ (black diamonds). b) and c) show $V_{\text{eff}}$
for $M=0.1$ and $M=0.01$, respectively. For clarity, curves representing
different values of $m$ are shifted relatively. }
 \label{fig:landscape}
\end{figure}

Computer simulation data for the effective potential $V_{\text{eff}}(y)$
are presented in Fig. \ref{fig:landscape}.
Two possibilities occur: if the $A$-particle is much heavier than the
$B$-particles, the potential is fully attractive by a combination of
depletion attraction close to the wall and gravity. In the brazil-nut
case, on the other hand, there are three regimes in $V_{\text{eff}}(y)$:
an attractive regime close to the wall (corresponding to situation I,II in
Fig. \ref{fig:bottom_jump}), followed by a repulsive regime (situation III
in Fig. \ref{fig:bottom_jump}) and a subsequent attractive regime
(situation IV in Fig. \ref{fig:bottom_jump}). The repulsive regime is
caused by the lift force
also responsible for the brazil-nut effect.

Fig. \ref{fig:landscape}a includes the theoretical
prediction of $V(y)$ from Eqn. (\ref{eq:V_theo}) for one set of
parameters.
There is very good agreement if the renormalized value $R^{\prime}$ for the depletion
bubble radius is taken (diamonds) while the agreement deteriorates for the
low-density-expression $R$ (circles). This demonstrates that the
analytical expression (\ref{eq:V_theo}) incorporates the basic physics principles.

Between the short-ranged wall attraction and the lift regime, there is an
energetic barrier $\Delta E$ typically of the order of several thermal
energies $k_BT$. Comparing our MC simulation results to approximation
(\ref{eq:V_theo}) with renormalized bubble radius, the energetic barrier
height proves to be in fair agreement. For $M=0.1, m=0.5$, Eqn.
(\ref{eq:V_theo}) predicts $\Delta E = 8.6 k_BT$, while we obtain $\Delta
E = 7.8 k_BT$ from our MC simulation. 
Considering the parameter combination $M=0.3, m=0.9$, an energetic barrier height $\Delta E = 15.5
k_BT$ follows from Eqn. (\ref{eq:V_theo}), while MC simulation yields
$\Delta E = 9.8 k_BT$. Here, this large discrepancy can be attributed to the relatively
thin layer of $B$-particles close to the bottom wall.

\begin{figure}
 \centering
 \includegraphics[width = 1.0\linewidth]{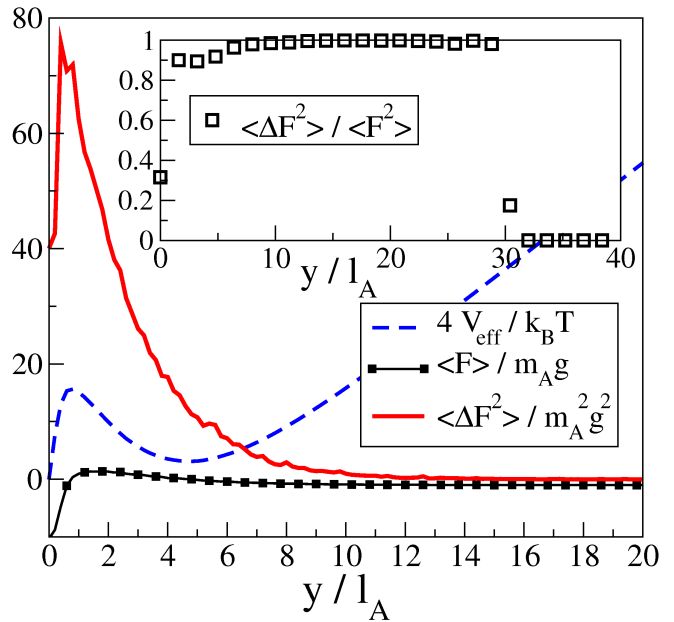}
 \caption{(Color online) Mean square fluctuation $\langle \Delta F^2
\rangle$ versus reduced height $y/l_A$. Inset: Relative fluctuation
$\langle \Delta F^2 \rangle / \langle F^2 \rangle$ versus reduced height
$y/l_A$. The parameters are $M = 0.1, m=0.5$.}
 \label{fig:var}
\end{figure}

If the energetic barrier is interpreted as a static one, an $A$-particle
which is initially trapped in the metastable minimum close to the wall
needs a huge escape time to leave this metastable minimum  which scales in
an Arrhenius-like fashion $\propto \exp \left( \Delta E /k_B T \right)$.
However,
as pointed out by Vliegenthart et al. in Ref.\ \cite{vliegenthart2003},
the effective interaction can strongly  fluctuate such that the energetic
barrier is not static. In this case, the particle will escape much more
quickly by waiting for a fluctuation which reduces the energetic barrier
instantaneously and accelerates the escape process.
Therefore, we have computed also the fluctuations of the depletion force
$\langle \Delta F^2 \rangle = \langle F^2 \rangle - \langle F \rangle^2$,
see Fig. \ref{fig:var}. Near the potential barrier, $\langle \Delta F^2
\rangle$ increases significantly. The inset of Fig. \ref{fig:var} shows
the relative fluctuations. They are indeed of the order one near the
effective potential maximum. This indicates that $A$-particles are exposed
to strongly fluctuating forces when attempting to escape the boundary
layer. Therefore, although the depletion potential comprises a potential
barrier in the order of several $k_BT$, particle transitions from the
boundary layer to higher altitudes  occur at much higher frequencies than
expected from the static activated Arrhenius expression $\propto \exp (
\Delta E /k_B T)$. In fact, this facilitates the boundary layer sampling
in the Monte-Carlo simulations of many $A$-particles to a large extent and
ensures sufficient equilibration.

\section{Results under time-dependent gravity (colloidal shaking)}
Finally, we turn to the nonequilibrium situation of time-dependent gravity,
see Eqn. \ref{eq:V_ext_{t}}, which is a simple model of colloidal shaking.
As far as our methods are concerned, we now use Brownian dynamics
simulations appropriate for colloids and dynamical density functional
theory. Various different starting configurations were used in the
simulations to obtain statistical averages which were all sampled from 
an interacting bulk system. This corresponds to an initial
homogeneous density field in DDFT. The swap fraction $\vartheta$ is
chosen to be $1/4$, i.e. we consider the case
that the time-average of the gravity is non-zero. In particular, we
discuss the emergence of a steady state
upon time-periodic gravity.

\begin{figure}[h]
 \centering
 \includegraphics[width = 1.0\linewidth]{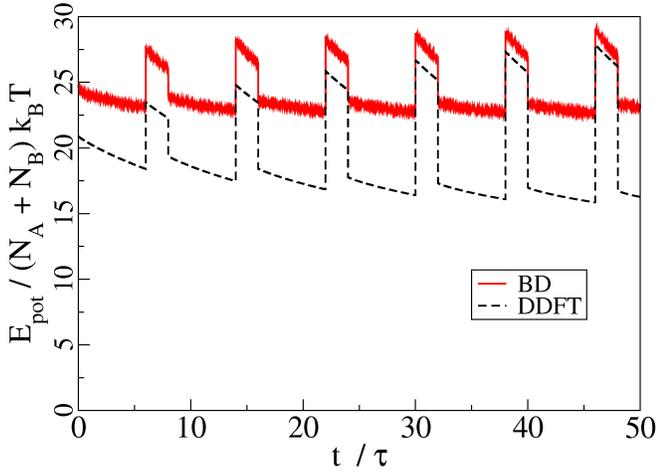}
 \caption{Total potential energy $E_{\text{pot}}/(N_A + N_B)$ per particle
versus reduced time $t/\tau$, using time period $T_0 = 8 \tau$ and swap
fraction $\vartheta = 1/4$. The inset shows the short time behavior. The
parameters are $M = 0.1, m = 0.24$.}
 \label{fig:bd_energy}
\end{figure}

The relaxation of an initially homogeneous (but interacting) fluid of $A$- and $B$-particles
towards its periodic steady  state can be  monitored by observing the
instantaneous ensemble-averaged total potential energy $E_{\text{pot}}$ of the
system \cite{assoud2009}:
\begin{align}
 E_{\text{pot}}(t) = &\frac{1}{2}\int \mathrm{d}^2r \hspace{-0.1cm} \int \mathrm{d}^2r' \hspace{-0.2cm}  \sum_{i,j = A,B} 
\hspace{-0.2cm}  u_{ij}(\bm{r} - \bm{r'}) \rho_i(\bm{r},t)\rho_j(\bm{r'},t)\nonumber \\
&+ \sum_{i = A,B}\int \mathrm{d}^2r V_{\text{ext},i}(\bm{r},t)\rho_i(\bm{r},t)
\end{align}
 This quantity is shown in Fig.
\ref{fig:bd_energy}, indicating that only few oscillations are needed to
get into the steady behavior. Due to the homogeneous starting configuration the energy oscillation amplitude increases with time.
DDFT describes all trends correctly and also provides good data for the potential energies and the associated
relaxation time.
 
\begin{figure}[ht]
 \centering
 \includegraphics[width = 1.0\linewidth]{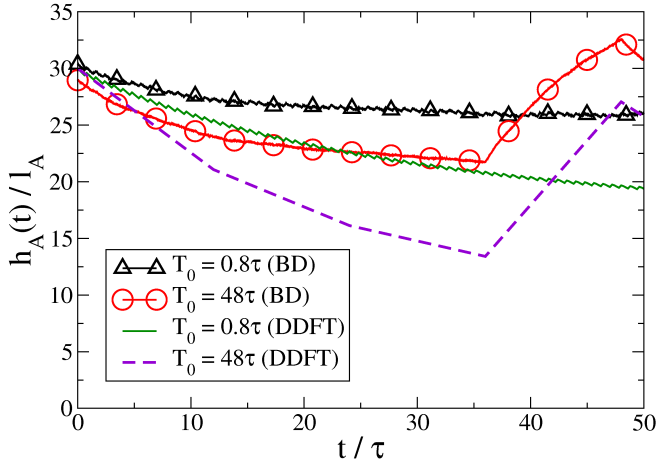}
 \caption{(Color online) Time-evolution of the mean $A$-particle height $h_A(t)$ versus
reduced time $t/\tau$ for different shaking periods $T_0$ and a swap fraction $\vartheta =
1/4$. The parameters are $M = 0.1, m = 0.24$.}
 \label{fig:bd_com}
\end{figure}

The averaged height, as defined by the first moment of the density profile (see Eqn. \eqref{eq:h}),
can be generalized to a dynamical (time-dependent) quantity $h_i(t), \ (i = A,B)$ via
\begin{equation}
\label{eq:bd_com}
h_i(t)= \frac{\int_0^{L_y} y \rho_i(y,t) \mathrm{d}y}{\int_0^{L_y} \rho_i(y,t)
\mathrm{d}y}, \qquad i = A,B \ .
\end{equation}
The time-dependent heights are another indicative parameter which probes the dynamical response
of the whole system \cite{rex2008,rex2009}. The quantity $h_A(t)$ is shown for two shaking
frequencies $\omega = 2\pi/T_0$ in Fig. \ref{fig:bd_com}. As a result, the relaxation
time is mainly scaling with the Brownian time $\tau$ but is rather
insensitive to the periodicity $T_0$. DDFT reproduces the trend with respect to increasing 
the periodicity $T_0$ but underestimates the actual heights $h_A(t)$, consistent with what we found in the equilibrium case.

\begin{figure}
 \centering
 \includegraphics[width = 1.0\linewidth]{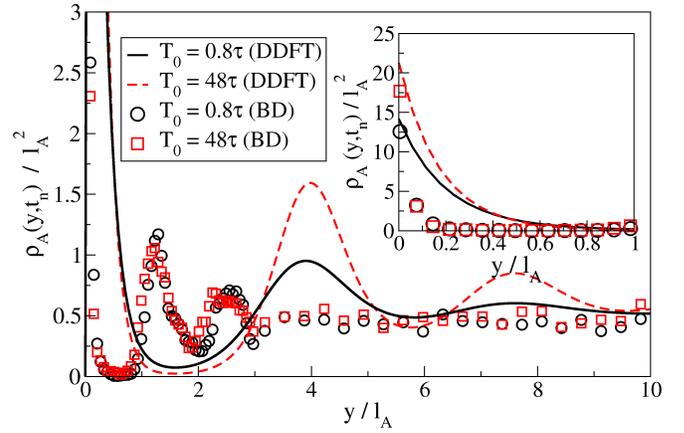}
 \caption{(Color online) $A$-particle density profile $\rho_A$ versus
reduced height $y/l_A$ upon shaking, obtained by DDFT (curves) and BD
simulations (symbols). The inset depicts the outer boundary peak
for both methods. Shaking
periods $T_{0}=0.8\tau$ and $T_{0}=48\tau$ are shown for a swap fraction
$\vartheta = 1/4$. The parameters are $M = 0.1, m=0.24$. }
 \label{fig:rho_dyn}
\end{figure}

Upon shaking, the boundary layer of the $A$-particles persists. We take
a snapshot after the relaxation time at
\begin{equation}
t_n = (n - \vartheta) T_0, \quad n = 1,2,..
\end{equation}
This is just the time at which the direction of gravity is inverted. 
 Comparing the results to the density
profiles predicted by DDFT, the persistence of the boundary layer is
indicated by both methods. Fig. \ref{fig:rho_dyn} depicts density profiles
obtained by BD simulations and DDFT for various shaking periods $T_0$.
Dynamical density functional data are in qualitative agreement with
Brownian dynamics computer simulations results but show the same 
deficiencies as in the equilibrium case. This indicates that the deviations 
can be solely attributed to the quality of the density functional but 
not to the additional adiabatic approximation inherent in any DDFT. 
Again, as in equilibrium, the amplitude of the outermost
density peak is in good agreement, see the inset of Fig. \ref{fig:rho_dyn}.

\begin{figure}[ht]
 \centering
 \includegraphics[width = 1.0\linewidth]{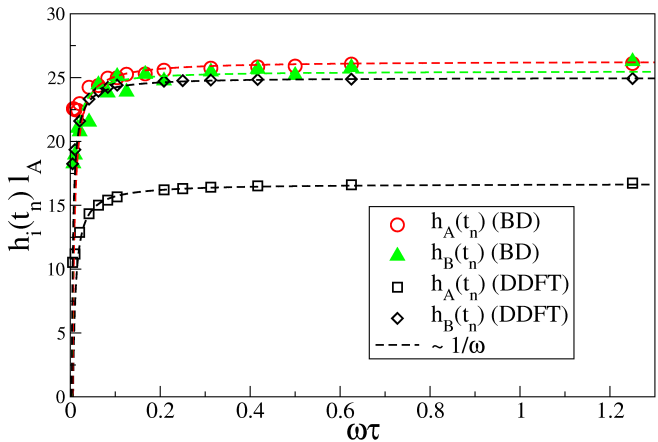}
 \caption{(Color online) Mean heights $h_i$ $(i=A,B)$ at times $t_n$ as a function of $\omega =
2\pi/T_0$ from DFT (squares and diamonds) and BD simulations (circles and triangles). The expected
scaling $\sim 1/\omega$ is indicated by dashed lines. The parameters are
$M =0.1, m=0.24$.}
 \label{fig:bd_w}
\end{figure}

The mean heights $h_i$ $(i=A,B)$ at times $t_n$ are shown in Fig. \ref{fig:bd_w} versus
frequency $\omega = 2\pi/T_0$.
The dissipative response leads to a decay of the peak as a function of
$\omega$. For $\omega \to 0$, we recover a quasi-static equilibrium case
while for
$\omega \to \infty$, the shaking is so fast that the system does not react
at all upon this stimulus. It therefore approaches the limit of an
ordinary equilibrium fluid mixture confined between two slits in the
absence of gravity where layering is much less pronounced than in the
presence of gravity. 
The crudest model of colloidal shaking is  a completely overdamped
response to a periodic external stimulus. In this case, the response
amplitude scales with the shaking frequency  as $1/\omega$. A fit which
involves the
$1/\omega$ scaling is included in Fig. \ref{fig:bd_w} and provides a good
description in the high-frequency regime. The DDFT reproduces these trends 
and provides good data for $h_B(t_{n})$ while underestimating $h_A(t_{n})$.
Again, we attribute this to the low-density approximation of the functional,  
where the stronger interacting $A$-particles are treated in a more 
approximative way than the $B$-particles. 

\section*{VI. CONCLUSIONS}

 In conclusion, we have explored a two-dimensional  binary mixture of
particles interacting via long-ranged repulsive forces
 in gravity by using computer simulation and density functional theory
theory. The
more repulsive $A$-particles create a depletion zone (void space) of less
repulsive $B$-particles around them
reminiscent to a bubble. Applying Archimedes' principle effectively to this
bubble, an $A$-particle can be lifted in a fluid background of
$B$-particles. This mechanism also work when the $A$-particles are heavier
than the $B$-particles leading to a colloidal brazil
nut effect where the heavier  particles float on top of the lighter 
particles. Still the buoyancy principle is fulfilled
if it is effectively applied to the mass per bubble volume.
This general finding is
in accordance with the recent experimental sedimentation results on
colloidal mixtures by Serrano et al. \cite{Serrano}.\\

Within the depletion bubble picture, an effective attraction of a
$A$-particles towards a hard container bottom wall is obtained
which leads to boundary layering of the $A$-particles. We have also studied
a periodic inversion of gravity causing
perpetuous mutual penetration of the mixture in a slit geometry.
This non-equilibrium case of time-dependent gravity is similar to shaking.
Upon shaking the boundary layering persists.
Our results are based on Brownian dynamics computer simulations and
density functional theory.
The brazil-nut and boundary layering are very general effects, they do
also occur for other
long-ranged repulsive interactions and in three spatial dimensions. They
are therefore verifiable in future settling  experiments on dipolar or
charged colloidal mixtures  as well
as in charged granulates and dusty plasmas.

Future work should address other interparticle interactions. Novel effects
are expected for a strongly attractive cross-interactions leading to
mutual mixing of $A$- and $B$-particles. These interactions are for
example realized in oppositely charged suspensions
\cite{leunissen2005,vissers2011}.
It would be interesting to check whether a colloidal brazil-nut effect can
still be observed in this case. Another option for future study is to
superimpose more external fields (e.g. an external electric field) to the
gravitational field
in order to controll the response of the system even more
\cite{hoffmann2000}. The latter set-up is relevant for electronic ink
\cite{comiskey1998, gelinck2004}.

\section*{Acknowledgments}

We thank M. Marechal for helpful discussions and L. Assoud for providing a
computer code.
This work was financially supported by the DFG within SFB TR6 (project C3).


\begin{thebibliography}{10}
\bibitem{footnote2}
The opposite case $m>1$ (for $0 \leq M \leq 1$) where the weaker coupled
  particles are heavier is not considered in this paper though it is in
  principle realizable.
\end{thebibliography}
\end{document}